\documentclass[useAMS]{mn2e}
\usepackage{graphicx}

\title[Gamma-ray Variability 3C 454.3]{Does the gamma-ray flux of the blazar 3C 454.3 vary on sub-hour time scales?}
\author[L. Foschini et al.]{L. Foschini\thanks{E-mail: \texttt{luigi.foschini@brera.inaf.it}.}, G. Tagliaferri, G. Ghisellini, G. Ghirlanda, F. Tavecchio, G. Bonnoli\\
INAF - Osservatorio Astronomico di Brera, via E. Bianchi 46, 23807, Merate (LC), Italy\\}

\begin{document}

\date{Accepted 2010 June 3.  Received 2010 May 18; in original form 2010 April 26}

\pagerange{\pageref{firstpage}--\pageref{lastpage}} \pubyear{2010}

\maketitle

\label{firstpage}

\begin{abstract}
In the early days of April 2010, the blazar 3C 454.3 ($z=0.859$) underwent a strong $\gamma-$ray outburst, reaching fluxes ($E>100$~MeV) in excess of $10^{-5}$~ph~cm$^{-2}$~s$^{-1}$. The \emph{Fermi Gamma ray Space Telescope} performed a $200$~ks long pointed observation starting from 5 April 2010 19:38 UTC. This allowed us to try probing the variability of the $\gamma-$ray emission on time scales of hours or less. We found the variability on a few hours time scale. On sub-hour time scale we found no evidence of significant variability, although the present statistics is not yet conclusive and further observations are needed. 
\end{abstract}


\section{Introduction}
Blazars are one type of active galactic nuclei (AGN) with relativistic jet. The small viewing angle of the jet makes it possible to observe strong effects of the special relativity, such as a boosting of the emitted power and a shortening of the characteristic time scales. Although the variability in blazars has been observed at all the frequencies (see, for example, the reviews by Ulrich et al. 1997 and Wagner 2008), the part at high energies deserves a specific interest, since most of the bolometric power of these sources is emitted at $\gamma$ rays.

The shortest time scales measured by the EGRET experiment onboard the \emph{Compton Gamma-Ray Observatory} are $\sim 4$ hours for PKS~1622-29 (Mattox et al. 1997) and $8$ hours for 3C~279 (Wherle et al. 1998). Early results from the Large Area Telescope (LAT) onboard the \emph{Fermi Gamma-ray Space Telescope} (hereafter \emph{Fermi}) indicated similar results: from about half-day for PKS~1454-354 (Abdo et al. 2009a) and PKS~1502+106 (Abdo et al. 2010a) to $5-6$ hours in the cases of 3C~454.3 and PKS~B1510-089 (Tavecchio et al. 2010) and 3C~273 (Abdo et al. 2010b).

These time scales are in agreement with the common paradigm that the characteristic spatial scale of the emitting region is of the order of the gravitational radius $r_{\rm g}=GM/c^2$ of the central spacetime singularity (Begelman et al. 2008). The zone where most of the dissipation occurs is located at distances greater than $\Gamma^2r_{\rm g}$, where $\Gamma$ is the bulk Lorentz factor (for most blazars $\Gamma \sim 10$, see Ghisellini et al. 2010). The size of the emitting region is in turn linked to the variability time $t_{\rm var}$ by the relationship $R < ct_{\rm var}\delta(1+z)$ (where $\delta$ is the Doppler factor). Obviously, it is expected that $R > r_{\rm g}$. Moreover, the emitting region must be optically thin and sufficiently far from the central black hole to allow the $\gamma$ rays to escape, otherwise they convert into electron-positron pairs.

An order-of-magnitude estimation of the above parameters results in an overall agreement with the observed variability on scales of days-hours. However, the recent detection by ground-based Cerenkov telescopes (HESS and MAGIC) of fast variability (minutes time scale) at $\gamma$ rays in some TeV BL Lac Objects severely threatens the above scenario. In 2006, during an exceptional outburst (average flux 7 Crab with peaks of more than 14 Crab), PKS 2155-304 displayed variability with doubling flux time scale of about $200$~s (Aharonian et al. 2007). In 2005, Mkn 501 changed its flux within a few minutes (Albert et al. 2007).  Given the masses of these two blazars of the order of $10^9M_{\odot}$, the measured variability is more than one order of magnitude smaller than the minimum allowed. Several solutions have been proposed, from a ``simple'' increase of the Doppler factor ($\delta > 50$) to invoking an internal structure of the jet (Begelman et al. 2008, Ghisellini \& Tavecchio 2008, Giannios et al. 2009).

As noted by Begelman et al. (2008), while these explanations could fit reasonably with BL Lac subclass of blazars, the same is not appropriate for flat-spectrum radio quasars (FSRQ) subclass. Indeed, this type of blazars generate $\gamma$ rays by inverse Compton on a population of seed photons external to the jet (external Compton, EC). Therefore, the pair photosphere can be at fairly large distances ($>10^4r_{\rm g}$), which in turn means that the blob has a size so large to determine an insufficient energy density to develop strong and fast flares. 

It is evident that the search for sub-hour time scales in FSRQ can have strong impact on the current knowledge. If measured, such very short variability would call for a strong revision of the present day models and understanding of relativistic jets. What we need is a strong outburst -- like those occurred in the BL Lac Objects PKS~2155-304 and Mkn~501 -- and a highly performing instrument.

To date, the shortest time scale observed at high energies in a FSRQ is $\sim 2000$~s, but in the hard X rays. It is the case of NRAO~530 as observed by \emph{INTEGRAL} ($20-40$~keV energy band) during an exceptional flare in February 2004 (Foschini et al. 2006). Despite of the good performance of \emph{INTEGRAL}, that episode consists of a single $5\sigma$ detection, while the source is below the detection limit of the instrument during the remaining of the time. No multiwavelength observations were possible at that time and, specifically, no $\gamma-$ray satellites were operating. Therefore, this remains an isolated exceptional case and, as for every exceptional claim without a strong observational support, there is always an ``halo'' of doubts.

The launch of \emph{Fermi}/LAT in June 2008 could give more opportunities to search for sub-hour variability. The Large Area Telescope (LAT, Atwood et al. 2009) could have the necessary sensitivity to probe the shortest time scales at $\gamma$ rays in quasars. If we consider a flux of $10^{-5}$~ph~cm$^{-2}$~s$^{-1}$ ($E>100$~MeV) and an effective area of $\sim 5000$~cm$^2$ (see Fig.~14 of Atwood et al. 2009, by taking into account the $0.1-100$~GeV energy range), it results that for normal incidence there are about $15$~photons per $5$~minutes (about a $4\sigma$ detection, by assuming that on such short time scales the background is almost absent). However, LAT operates almost always in scanning mode. This, in turn, has the great advantage to offer an efficient monitoring of all the sky (a complete coverage every three hours or two orbits), but it hampers the possibility to probe time scales shorter than three hours (the source is not always at the boresight of the instrument and, therefore, the effective area is smaller). 

A first tentative to bypass this issue has been recently made by performing a pointed observation during the giant outburst of 3C 454.3 ($z=0.859$) occurred in early April 2010 (Wallace et al. 2010), when the source reached fluxes in excess of $10^{-5}$~ph~cm$^{-2}$~s$^{-1}$ ($E>100$~MeV). Such flux levels were already reached by 3C 454.3 during the first ten days of December 2009 (see Bonnoli et al. 2010 and references therein), but it was not possible at that time to perform a pointed observation and hence to search for sub-hours time scales. 

\begin{figure*}
\centering
\includegraphics[angle=270,scale=0.35]{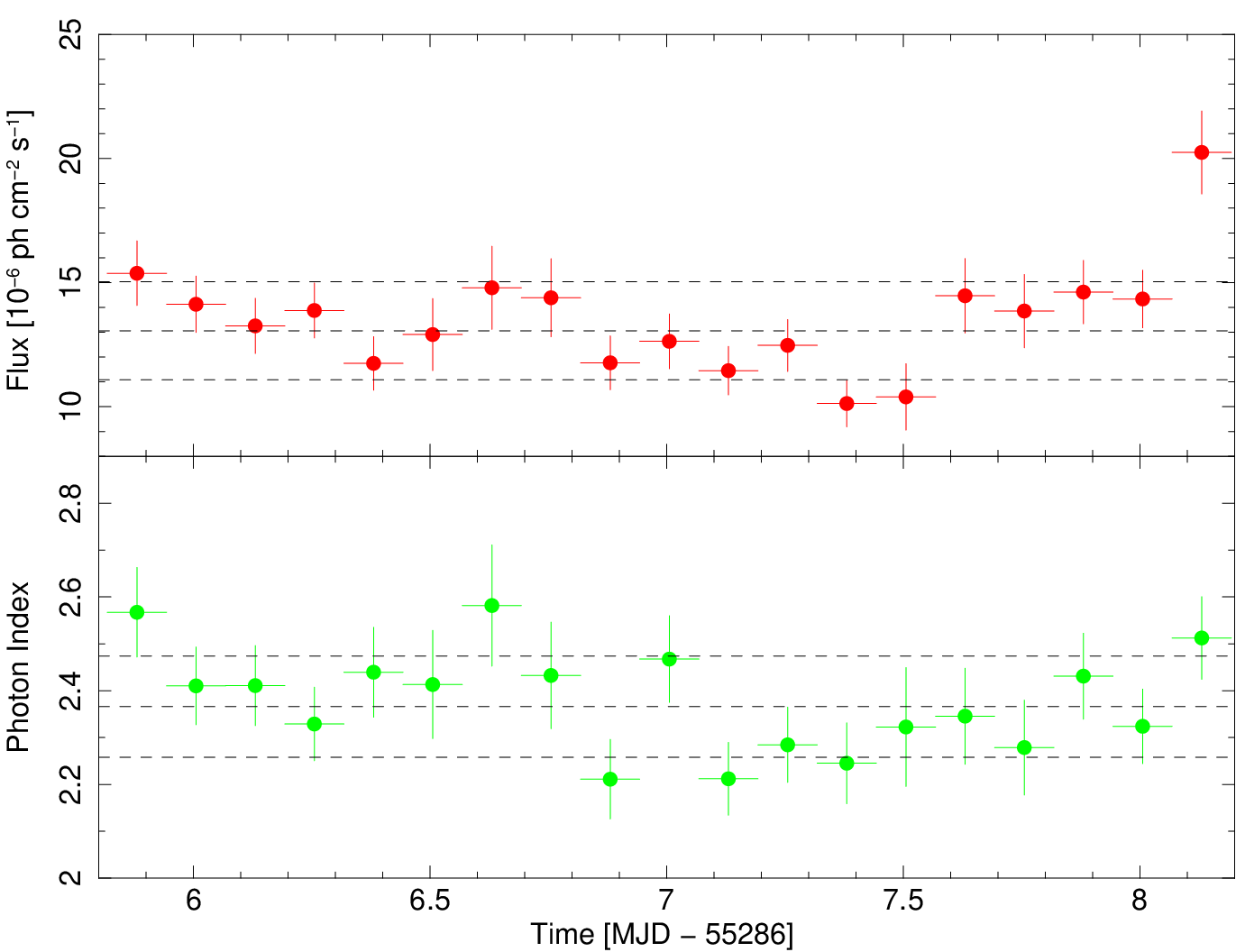}
\includegraphics[angle=270,scale=0.35]{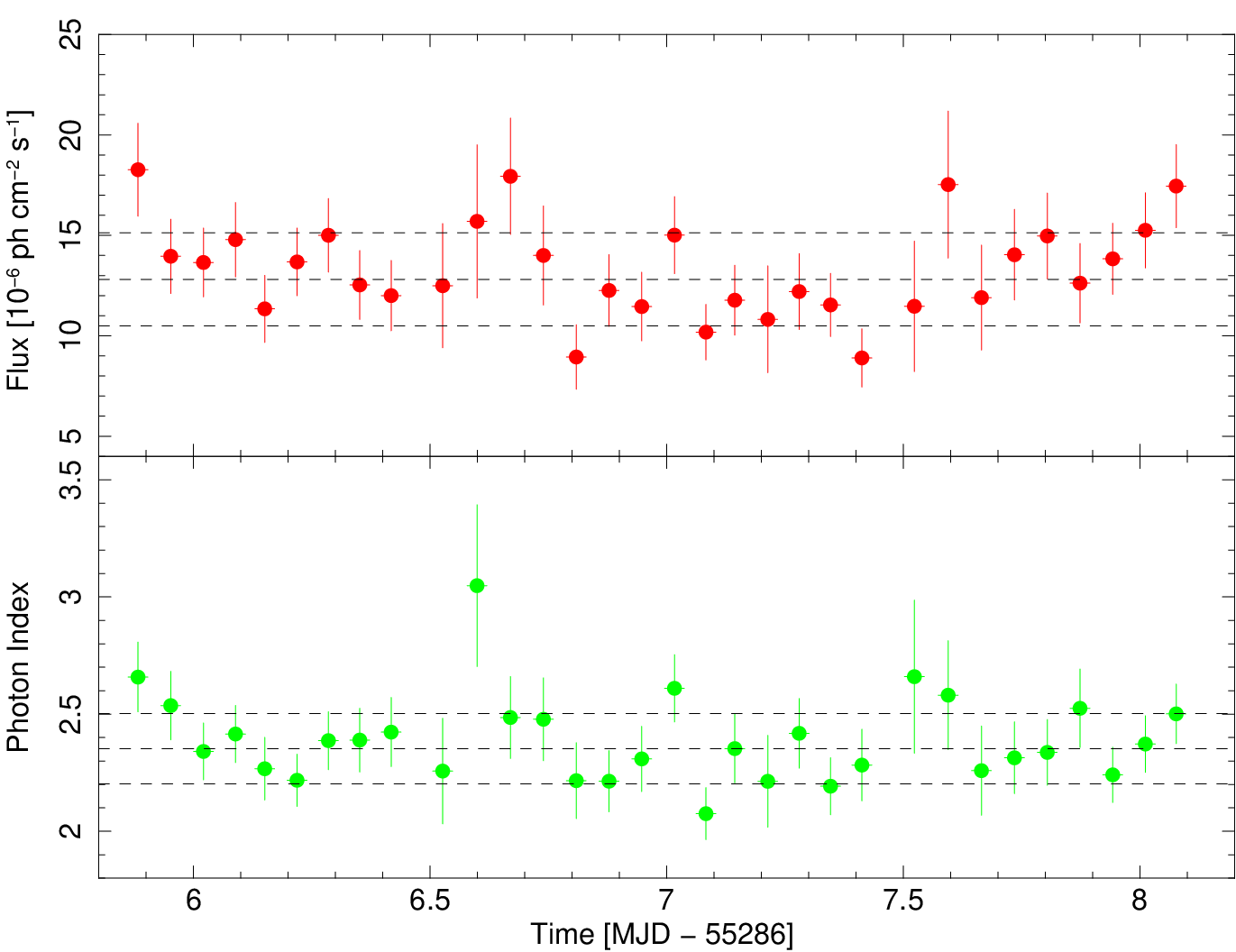}
\caption{Light curves of 3C 454.3 ($E>100$~MeV) with 3 hours \emph{(left panel)} and 1 hour \emph{(right panel)}, respectively. Time starts on 31 March 2010 (MJD 55286), so that the number of the days corresponds also to days of April. The horizontal dashed lines in each panel correspond to the weighted average plus/minus one standard deviation.}
\label{fig:curva3p1}
\end{figure*}

Here we report the study of the variability of the $\gamma-$ray emission of 3C 454.3 as observed by \emph{Fermi}/LAT during the pointed observation performed from 5 to 8 April 2010.

\section{Data Analysis}
Data of \emph{Fermi}/LAT (Atwood et al. 2009) have been downloaded from the publicly accessible web site of HEASARC\footnote{\texttt{http://fermi.gsfc.nasa.gov/}} and analyzed by means of the public software package \texttt{LAT Science Tools v. 9.15.2}, including the calibration files (Instrument Response File, IRF \texttt{P6\_V3\_DIFFUSE}), the maps of the spatial diffuse background and the spectrum of the isotropic (cosmic and instrumental) background. 

We retrieved the photon list with $E>100$~MeV and spacecraft data from 5 April 2010 19:38 UTC to 8 April 2010 03:12 UTC (MJD $55291.8180-55294.1333$) corresponding to an elapsed time of about $200$~ks (see Fig.~\ref{fig:curva3p1}). Events of class 3 (diffuse) inside a radius of $7^{\circ}$ centered on the coordinates of 3C 454.3 ($RA=22^{\rm h}:53^{\rm m}:57^{\rm s}.7$; $Dec=+16^{\circ}:08':53''$, J2000) and coming within $105^{\circ}$ of the zenith angle have been selected. This data set has been further filtered to exclude the events outside the good-time intervals and within $8^\circ$ of the Earth limb. Livetime and exposure maps have been calculated, as well as the diffuse background.

An unbinned likelihood algorithm (Cash 1979, Mattox et al. 1996), implemented in the \texttt{gtlike} task, has been used to extract the flux and the photon index of the source emission. A single power-law model in the form $F(E)\propto E^{-\Gamma}$, where $\Gamma$ is the photon index, was used to fit the energy flux distribution of the blazar. It is known that the $\gamma-$ray spectrum of 3C 454.3 displays a break at a few GeV (Abdo et al. 2009b, Finke \& Dermer 2010), but since we want to probe the shortest time scales (hence, lower statistics), it is better to adopt a single power-law model. The broken power-law model has been used to fit the data of the whole observation (see the next Section). The backgrounds (spatial diffuse and isotropic) measured during this fit are kept fixed during the processing of the light curves with hours time bins or less, since we do not expect strong changes of the background in such short time scales. 

A first run with the optimizer \texttt{DRMNFB} is done to grossly estimate the spectral parameters. Then, the output of this process is used as input for a second run with the \texttt{NEWMINUIT} optimizer, to better assess the measurement errors. In the following, only statistical errors are displayed. The use of the \texttt{P6\_V3\_DIFFUSE} response file guarantees that systematics affect the results by a value of less than 20\% (Rando et al. 2009). However, since this work is based only on LAT data, with no comparison with other instruments at other wavelengths, there is no need of absolute flux calibration. Hence, systematics are not applied. 

This procedure is performed iteratively for each bin time within each good-time interval in order to generate the light curve. We discarded all the detections that have no sufficiently high \emph{Test Statistic} ($TS$, see Mattox et al. 1996 for the definition), i.e. we considered only time bins with detection at $\sigma \sim \sqrt{TS}>5$. A light curve with 3 hours time bin is displayed in Fig.~\ref{fig:curva3p1}, left panel. The effective source ontime changes within the elapsed time bin, ranging from 10 to 98\% in the 30-minutes bin light curve. Therefore, we checked for a reasonable high $TS$ even in presence of low source ontime. For example, the bin with the lowest source ontime (10\% of the elapsed time bin) has $TS = 73$.

We note that the average $TS$ increased by a factor $\approx 2.5$ during the pointed observation, as expected from the increase of the instrument sensitivity when changing from scanning to pointed mode (the light curve in scanning mode is not shown).

\section{Discussion}
The whole data set of $200$~ks observation has been fitted with a broken power-law model in the form of $F(E)\propto (E/E_{\rm break})^{\Gamma_1}$ for $E<E_{\rm break}$ and $F(E)\propto (E/E_{\rm break})^{\Gamma_2}$ otherwise. This results in these parameters: the photon indices below and above the break energy are $\Gamma_1=2.22\pm 0.04$ and $\Gamma_2=2.80\pm 0.08$, respectively. The break energy is $0.9\pm 0.2$~GeV and the flux integrated over the $0.1-100$~GeV energy band is $(12.7\pm 0.3)\times 10^{-6}$~ph~cm$^{-2}$~s$^{-1}$. The recorded counts are $3044$ and $TS=10592$ ($>100\sigma$). We compare these values with the early observation with \emph{Fermi} referring to the first month of data (Abdo et al. 2009b). In August 2008, the flux was about a factor 4 smaller than in April 2010, with a value of $(3.0\pm 0.1)\times 10^{-6}$~ph~cm$^{-2}$~s$^{-1}$, the break energy was greater ($2.4\pm 0.3$~GeV), and the photon index above the break quite softer, with a value of $\Gamma_2=3.5\pm 0.2$. $\Gamma_1$ is consistent within the measurement errors. It seems that the higher flux needs of a harder spectrum.

The $\gamma-$ray emission of 3C 454.3 in Fig.~\ref{fig:curva3p1} (left panel) is variable on a few hours time scale. A fit with a constant-flux line gave $\chi^2/dof=\tilde{\chi}^2\sim 2.7$ (with 18 degrees of freedom, this means a chance probability of 0.99). However, we note that the last point, which is mainly driving the global variability of the light curve, is not a statistical fluctuation, but the beginning of a real flux increase. This is confirmed by the fact that on 8 April there was the peak of the $\gamma-$ray emission of the April 2010 outburst, with a day average of $\sim 16\times 10^{-6}$~ph~cm$^{-2}$~s$^{-1}$. On the other hand, no strong variability of the photon index has been measured, although some trend is visible between $\sim 6.5$ and $\sim 7$ April: $\Gamma$ changed from $2.58\pm 0.13$ on $\sim 6.63$~April to $2.21\pm 0.08$ ($\sim 3\sigma$ significance) on $\sim 6.88$~April. Similar spectral changes have been observed in $\gamma-$ray blazars by EGRET (Nandikotkur et al. 2007) and LAT in the case of 3C 273 (Abdo et al. 2010b). 

\begin{figure}
\centering
\includegraphics[angle=270,scale=0.35]{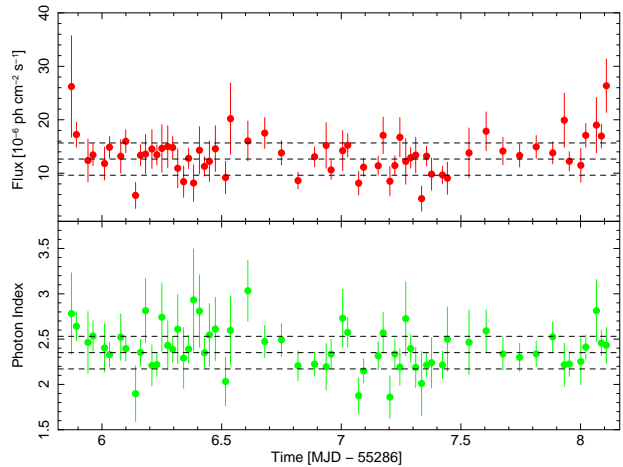}
\caption{\emph{(top panel)} Flux light curve of 3C 454.3 ($E>100$~MeV) with 30 minutes time bin; \emph{(bottom panel)} photon index. Time starts on 31 March 2010 (MJD 55286), so that the number of the days corresponds also to days of April. The horizontal dashed lines in each panel correspond to the weighted average plus/minus one standard deviation.}
\label{fig:curva30min}
\end{figure}

The light curve with 1 hour time bin (Fig.~\ref{fig:curva3p1}, right panel) has larger error bars and hence the slight variability seen in the 3-hours-bin curve is now smoothed. The fit with a constant flux line gives $\tilde{\chi}^2=1.6$, but the visible trends are consistent with those observed in Fig.~\ref{fig:curva3p1} (left panel). The most significant trend is the drop in flux observed between $\sim 6.7-6.8$~April.

To evaluate the time scales of these trends, we calculated the time of exponential rise or decay as defined by:

\begin{equation}
F(t) = F(t_0)\exp[-(t-t_0)/\tau]
\end{equation}

where $F(t)$ and $F(t_0)$ are the fluxes at the time $t$ and $t_0$, respectively, and $\tau$ is the characteristic time scale. 

In this case, the drop on $\sim 6.7-6.8$~April visible in Fig.~\ref{fig:curva3p1} (right panel), has $\tau = 4.8\pm 4.6$ hours (significance of the flux variation $3.1\sigma$). This value is confirmed in the 30-minutes time bin light curve shown in Fig.~\ref{fig:curva30min} ($\tau = 4.7\pm 4.4$~hrs, $3.1\sigma$). Moreover, the rise at the end of the curve between $\sim 8.0-8.1$ has $\tau = 2.7\pm 1.0$~hrs ($3.3\sigma$). 

It is possible to estimate the minimum Doppler factor $\delta$ (e.g. Dondi \& Ghisellini 1995, Mattox et al. 1997)\footnote{In the present work we used the most recent value for the Hubble-Lema\^{i}tre constant $H_0=73$~km~s$^{-1}$~Mpc$^{-1}$ (Freedman \& Madore 2010).}, although during this pointed observation it was not possible to perform multiwavelength observations, because the blazar was too close to the Sun. Therefore, we adopted for the X-ray flux and spectrum, the values measured during the December 2009 outburst, when 3C~454.3 reached similar $\gamma-$ray fluxes: $\alpha=0.4$ and $F_{\rm 1~keV}\sim 30$~$\mu$Jy (Bonnoli et al. 2010). We obtain $\delta \geq 14$, for 1~GeV $\gamma$ rays and $\tau = 2.7$~hours, a high value for a lower limit, but not unlikely (cf Ghisellini et al. 2010).

We noted a few cases of sudden drop in the flux with $\sim 30$~min time scale on $\sim 6.1-6.2$ and $\sim 7.3$~April. A closer inspection of the data revealed that these bins had small source on time ($\sim 25$\% of the whole bin). Therefore, it is likely that it is a fake drop, caused by the not good reconstruction of the flux in presence of too few events, although sufficient for a high $TS$.

\section{Final remarks}
We presented the variability analysis of the $\gamma-$ray observation with \emph{Fermi}/LAT of the blazar 3C 454.3 pointed during the outburst of early April 2010. Contrary to the usual scanning mode, this time \emph{Fermi} performed a $200$~ks pointed observation (Wallace et al. 2010), which guarantees a higher sensitivity and a longer source ontime. These particular settings, together with the exceptionally high flux emitted by the blazar, enabled us to probe the shortest time scales that we found to be in the range of $\sim 2-5$ hours, confirming the indications found by Bonnoli et al. (2010), Tavecchio et al. (2010) and in agreement with the current theories.

Although the significance of the events is not very strong ($\sim 3\sigma$), the fact that it is present in the light curves with different time bins is an indication that the findings are well grounded. On the other side, we have not found significant evidence of sub-hour time scales, but the present data set is not yet conclusive. We noted that the flux increased at the end of the pointed observations with $\tau\sim 2.7$ hours. Therefore, we could expect (or reasonably hope) that hours or less time scales may be effectively probed with \emph{Fermi}/LAT under conditions of even greater fluxes, likely as occurred in December 2009 (Bonnoli et al. 2010).

\section*{Acknowledgments}
This research has made use of data obtained from the High Energy Astrophysics Science Archive Research Center (HEASARC), provided by NASA's Goddard Space Flight Center. We are grateful to the \emph{Fermi}/LAT Collaboration for having performed this pointed observation and made the data publicly available immediately. This work has been partially supported by PRIN-MiUR 2007 and ASI Grant I/088/06/0.

\label{lastpage}

\end{document}